\RequirePackage{etoolbox}
\csdef{input@path}{{style/}{graphics/}}
\documentclass[aoas,MSNbibl,nameyear,dvips]{arximspdf}
\makeatletter
   \@ifpackageloaded{graphicx}{}{\usepackage{graphicx}}
\makeatother
\usepackage{dcolumn}
\usepackage{multirow}
\usepackage{graphicx}
\usepackage{url,breakurl}

\doi{10.1214/14-AOAS793} 
\volume{9}
\issue{1}
\pubyear{2015}
\firstpage{43}
\lastpage{68}
\docsubty{FLA}

\makeatletter
\newcolumntype{d}[1]{D{.}{.}{#1}}
\renewcommand{\mid}{|}
\newcommand{\pkg}[1]{{\fontseries{b}\selectfont #1}}

\newcommand{\bxi}{\bolds{\xi}}

\newcommand{\bgamma}{\bolds{\gamma}}
\newcommand{\blambda}{\bolds{\lambda}}
\newcommand{\bbeta}{\bolds{\beta}}

\newcommand{\bOmega}{\bolds{\Omega}}

\def\calD{\mathcal{D}}

\def\calL{\mathcal{L}}

\def\CPO{\mathrm{CPO}}
\def\LPML{\mathrm{LPML}}

\def\DIC{\mathrm{DIC}}

\newcommand{\ba}{\mathbf{a}}

\newcommand{\be}{\mathbf{e}}

\newcommand{\bmmm}{\mathbf{m}}

\newcommand{\bI}{\mathbf{I}}

\newcommand{\bS}{\mathbf{S}}

\newcommand{\bw}{\mathbf{w}}

\newcommand{\bx}{\mathbf{x}}

\newcommand{\by}{\mathbf{y}}

\newcommand{\bz}{\mathbf{z}}

\newcommand{\boeta}{\bolds{\eta}}

\makeatother

\begin{document}
\begin{frontmatter}

\title{Modeling county level breast cancer survival data using a
covariate-adjusted frailty proportional hazards model}
\runtitle{Covariate-adjusted frailty PH model}

\begin{aug}
\author[A]{\fnms{Haiming}~\snm{Zhou}\thanksref{M1,T1}},
\author[A]{\fnms{Timothy} \snm{Hanson}\corref{}\thanksref{M1,T1}\ead[label=e2]{hansont@stat.sc.edu}},
\author[B]{\fnms{Alejandro} \snm{Jara}\thanksref{M2,T2}}\\
\and
\author[C]{\fnms{Jiajia} \snm{Zhang}\thanksref{M1,T1}}
\runauthor{Zhou, Hanson, Jara and Zhang}
\affiliation{University of South Carolina\thanksmark{M1} and Pontificia
Universidad Cat\'olica de Chile\thanksmark{M2}}
\address[A]{H. Zhou\\
T. Hanson\\
Department of Statistics\\
University of South Carolina\\
Columbia, South Carolina 29208\\
USA\\
\printead{e2}}
\address[B]{A. Jara\\
Department of Statistics\\
Pontificia Universidad Cat\'{o}lica de Chile\\
Santiago\\
Chile}
\address[C]{J. Zhang\\
Department of Epidemiology and Biostatistics\\
University of South Carolina\\
Columbia, South Carolina 29208\\
USA}
\end{aug}
\thankstext{T1}{Supported by NCI Grants R03CA165110 and 5R03CA176739,
and ASPIRE grant from the University of South Carolina.}
\thankstext{T2}{Supported by Fondecyt Grant 1141193.}

\received{\smonth{7} \syear{2014}}
\revised{\smonth{10} \syear{2014}}

%
\begin{abstract}
Understanding the factors that explain differences in survival times is
an important issue for establishing policies to improve national health systems.
Motivated by breast cancer data arising from the Surveillance
Epidemiology and End Results program, we
propose a covariate-adjusted proportional hazards frailty model for the
analysis of clustered right-censored data. Rather than incorporating
exchangeable frailties in the linear predictor of commonly-used
survival models, we allow the frailty distribution to flexibly change
with both continuous and categorical cluster-level covariates and model
them using a dependent Bayesian nonparametric model. The resulting
process is flexible and easy to fit using an existing R package. The
application of the model to our motivating example showed that,
contrary to intuition, those diagnosed during a period of time in the
1990s in more rural and less affluent Iowan counties survived breast
cancer better. Additional analyses showed the opposite trend for
earlier time windows. We conjecture that this anomaly has to be due to
increased hormone replacement therapy treatments prescribed to more
urban and affluent subpopulations.
\end{abstract}

%
\end{frontmatter}

\section{Introduction}\label{sec1}

Based on data gathered for Iowa State in the Surveillance Epidemiology
and End Results (SEER) program of the National Cancer Institute, we
assess the effect of potential risk factors for womens' breast cancer.
This involves the analysis of clustered time-to-event right-censored
data, where event times of patients from the same county of residence
are expected to be associated with each other, possibly due to sharing
common unobserved characteristics, such as region-specific differences
in environments, treatment resources or diagnosis of the patients. As
is widely known, taking into account the clustered nature of the data
is a must to obtain valid statistical inferences [see, e.g., \citet
{Therneau2000}, Chapter~8].

A standard way of modeling clustered survival data is to introduce a
common random effect (frailty) into the survival model for each
cluster, yielding shared frailty models. ``Frailties,'' termed by \citet
{Vaupeletal1979}, were originally introduced to deal with possible
heterogeneity due to unobserved covariates and are regarded as
unobserved common characteristics for each cluster able to account for
the dependence among event times. In the context of the proportional
hazards (PH) model, as conventionally implemented, frailties are
incorporated into the linear predictor, and the median or mean of the
frailty distribution is constrained to be zero to avoid identifiability
problems. Conditional on the frailty, the model retains its
interpretation in terms of constants of proportionality of the hazards.
Survival models with frailties have been extensively used in the
statistical literature, especially when the comparison of event times
within cluster is of interest.

A common assumption in shared frailty survival models is the one of
homogeneity, where the frailties are assumed to be independent and
identically distributed (i.i.d.) random variables from a parametric or
nonparametric distribution [see, e.g., \citet
{ClaytonCuzick1985,Gustafson1997,Qiouetal1999,WalkerMallick1997}].
Although the nonparametric approach provides flexibility in capturing a
frailty distribution's variance, skewness, shape and even modality, it
essentially assumes that these frailty distributional aspects are the
same across all the clusters, which may be restrictive for particular
data sets [\citet{Nohetal2006}]. For example, in the kidney
transplantation study, \citet{Liuetal2011} argue that the frailty
distribution may be affected by some cluster-level covariates, since
``\ldots\emph{urban transplant facilities may exhibit more uniform
practices than rural transplant hospitals, corresponding to less
heterogeneity} (\emph{smaller variance}) \emph{for frailties of urban centers}\ldots'' Ignoring such heterogeneity can drastically affect the inference
for cluster-specific effects and prediction [\citet{McCullochNeuhaus2011}].

As the process generating the frailty terms is on its own right of
scientific interest, different extensions of the i.i.d. frailty
modeling approach have been considered. \citet{WassellMoeschberger1993}
studied the impact of interventions in the Framingham Heart Study by
introducing a modified gamma frailty with a pairwise
covariate-dependent parameter. \citet{YashinIachine1999} considered the
dependence between frailty and observed covariates (BMI and smoking) in
Danish twins to investigate the heritability of susceptibility to
death. \citet{Nohetal2006} verified frailty distribution heterogeneity
in a well-known kidney infection data set by applying a dispersed
normal model. \citet{Cottone2008} assumed either Bernoulli or normal
distributions for the frailties where the frailty distribution mean or
variance depends on cluster-level covariates through specified link
functions. \citet{Liuetal2011} proposed a covariate-dependent positive
stable shared frailty model with an application to kidney
transplantation data from the Scientific Registry of Transplant
Recipients, and demonstrated the heterogeneity in facility performance.
\citet{WangLouis2004} studied a related approach for binary data that
has both conditional and marginal interpretation using the so-called
bridge distribution instead of positive stable.

The previously described model extensions allow for particular and
specific aspects of distributional shape to change with cluster-level
covariates. However, a~more thorough evaluation of the effect of the
predictors should account for potential changes in characteristics of
the frailty distribution other than just, for example, the location or
scale. It is, for instance, useful to examine potential changes in the
skewness, symmetry and multimodality of the frailty distribution.
Therefore, a nonparametric formulation that anticipates changes in
shape, skew and modality beyond simple location models is of interest.

In this paper, we propose a practicable and general framework for
modeling clustered survival data as a function of covariates, based on
a predictor-dependent Bayesian nonparametric model for the frailties
and the Cox's PH model. Under the proposed approach the frailty
distribution flexibly changes with both continuous and categorical
cluster-level covariates, thus allowing for full heterogeneity across
clusters. We apply this modeling approach to a subset of the SEER
county-level breast cancer data consisting of 1073 women diagnosed with
malignant breast cancer during 1995--1998. Important patient-level
covariates include age at diagnosis, race, county of residence and the
stage of the disease. Additional county-level covariates potentially
associated with breast cancer survival are also available from census
data, including median household income, poverty level, education and a
rurality measure. These area-level socioeconomic factors have been
discovered to be associated with breast cancer by many researchers
[e.g., \citet{Spragueetal2011}]. Women living in more affluent or less
rural geographic areas tend to survive breast cancer better after a
diagnosis than those living in regions with indicators of low
socioeconomic status. Moreover, rural counties may present more
heterogeneity in access to quality care and screening for breast
cancer, leading to more variability for frailties of rural counties
[\citet{ZhaoHanson2011}]. This suggests to us that the frailty
distribution could be potentially affected by these county-level
socioeconomic factors. The results show that the proposed model
provides better goodness of fit to the data and is predictively
superior to the traditional PH spatial frailty model, as well as
helping to piece together a plausible story for the data in terms of
the prescribing of hormone replacement therapy.

The paper is organized as follows. In Section~\ref{sec2} we introduce the
proposed frailty PH model, including a detailed description of the
dependent Bayesian nonparametric model and the Markov chain Monte Carlo
(MCMC) implementation of the posterior computations. Section~\ref{sec3} provides
a detailed analysis of the motivating data set. Section~\ref{sec4} presents the
results of simulation studies to evaluate the performance of the
proposed model. Some concluding remarks and a final discussion are
given in Section~\ref{sec5}.

\section{Covariate-adjusted frailty proportional hazards model}\label{sec2}
\subsection{The modeling approach}\label{sec2.1}
Suppose that right-censored survival data $(\bw_{ij}, t_{ij}, \delta
_{ij})$ are collected for the $j$th subject of the $i$th cluster, where
$j=1,\ldots,n_i$, $i=1,\ldots,n$, $\bw_{ij}$ is a $p$-dimensional
vector of exogenous covariates, $t_{ij}$ is the recorded event time,
and $\delta_{ij}$ is the censoring indicator equaling $1$ if $t_{ij}$
is an observed event time and equaling $0$ if the event time is
right-censored at $t_{ij}$. Let $T_{ij}$ and $C_{ij}$ be the event and
censoring times, respectively, for the $j$th subject in the $i$th
cluster. To take into account the within-cluster association structure,
a~frailty PH model is assumed for $T_{ij}$. The conditional PH
assumption implies that the hazard function of $T_{ij}$ is given by
%
\begin{equation}
\label{frailtyPH} \lambda(t | \bw_{ij}, e_i) =
\lambda_0(t) \exp\bigl(\bw'_{ij}\bxi+
e_i\bigr),
\end{equation}
where $\be=(e_1, \ldots, e_n)'$ is an unobserved vector of frailties,
and $\lambda_0(t)$ is the baseline hazard function corresponding to the
event time of a subject with covariates $\bw=\mathbf{0}$ and $e=0$. We
additionally assume a conditionally independent censoring scheme, that
is, $C_{ij}$ and $T_{ij}$ are independent given $\bw_{ij}$ and $e_i$.
Often the frailties are assumed to be exchangeable or i.i.d. from some
parametric or nonparametric distribution $G$. For instance, \citet
{Therneauetal2003} considered exchangeable Gaussian frailties and
proposed an estimation procedure based on a Laplace approximation of
the likelihood function leading to a penalized partial likelihood. This
approach, referred to below as GF, will be compared with our method in
the simulation studies.

Now consider a partition of the predictor vector $\bw_{ij} = (\tilde{\bw
}'_{ij}, \bx'_i)'$, where $\bx_i \in\mathcal{X} \subseteq\mathbf
{R}^q$ is a $q$-dimensional vector of cluster-level covariates and
$\tilde{\bw}_{ij}$
is a $(p-q)$-dimensional vector of subject-specific covariates,
respectively, and the corresponding partition of the regression
coefficient vector $\bxi=(\tilde{\bxi}{}', \bxi'_x)'$. On the scale of
the linear predictor $\bw'_{ij}\bxi+ e_i$, the frailty $e_i$ models
the cluster-specific behavior and its distribution $G$ is shifted by
$\bx'_i\bxi_x$. Therefore, the homogeneity assumption implies that the
vector of cluster-level covariates $\bx_i$ modifies only the location
of the distribution of cluster-specific effects but not its shape. To
relax this assumption, we consider a covariate-adjusted frailty PH
model, where the frailty distribution depends on cluster-level
covariates $\bx_i$. That is,
\[
e_i \mid G_{\bx_i} \stackrel{\mathrm{ind}.}{\sim} G_{\bx_i},
\]
where for every $\bx\in\mathcal{X}$, $G_{\bx}$ is a probability
measure defined on $\mathbb{R}$; this specifies a probability model for
the entire collection of probability measures $\mathcal{G}^{\mathcal
{X}}=\{G_{\bx}\dvtx \bx\in\mathcal{X}\}$, such that its elements are
allowed to smoothly vary with the cluster-level covariates $\bx$.
Specifically, we consider
a mixture of linear dependent tailfree processes (LDTFP) prior [\citet
{JaraHanson2011}] for $\mathcal{G}^{\mathcal{X}}$, denoted as
\[
\mathcal{G}^{\mathcal{X}} \mid J,h,\theta, c,\rho\sim\operatorname{LDTFP}\bigl(h,\Pi^{J,\theta},\mathcal{A}^{J,c,\rho}\bigr),
\]
and
\[
c \mid Q \sim Q,
\]
where $J \in\mathbb{N}$ is the level of specification of the process,
$c \in\mathbb{R}_+$ is a prior precision parameter controlling the
prior variability of the process, $h(\cdot)= \frac{\exp\{\cdot\}}{1+\exp
\{\cdot\}}$, $\Pi^{J,\theta}$ is a $J$-level sequence of binary
partitions of $\mathbb{R}$, depending on the scale parameter $\theta
\in\mathbb{R}_+$,
$\mathcal{A}^{J,c,\rho} =\{ 2n/c\rho(1),\ldots, 2n/c\rho(J)\}$ is a
collection of positive numbers depending on $J$, $c$ and $\rho$, $\rho
\dvtx \mathbb{N} \longrightarrow\mathbb{R}_+$ is an increasing
function, and $Q$ is a probability measure defined on $\mathbb{R}_+$.

The LDTFP is specified such that for every $\bx\in\mathcal{X}$, the
process $G_{\bx}$ is centered around an $N(0,\theta)$ distribution,
that is, $E(G_{\bx})=N(0,\theta)$, for every $\bx\in\mathcal{X}$.
Furthermore, the process is specified such that for every $\bx\in
\mathcal{X}$, $G_{\bx}$ is almost surely a median-zero probability
measure. The latter property is important to avoid identifiability
problems. The LDTFP process includes as important special cases a
nonparametric exchangeable frailty model where $G_{\bx} = G_{\bx'}$ for
$\bx' \ne\bx$ as well as exchangeable normal frailties $G_{\bx
}=N(0,\theta)$ for all $\bx\in\mathcal{X}$.

As shown by \citet{JaraHanson2011}, dependent tailfree processes have
appealing theoretical properties, such as continuity as a function of
the predictors, large support on the space
of conditional density functions, straightforward posterior computation
relying on algorithms for fitting generalized linear models, and the
process closely matches conventional Polya tree priors [see, e.g., \citet
{Hanson2006a}] at each value of the predictor, which justify its choice
here. Polya trees have been extensively studied in the literature and
have desirable properties in terms of support and\vspace*{1pt} posterior
consistency. Details on the trajectories of $\operatorname{LDTFP}(h,\Pi
^{J,\theta},\mathcal{A}^{J,c,\rho})$, useful for a complete
implementation of algorithms for exploring the corresponding posterior
distributions, are given in Appendix~A of the supplementary material
[\citet{Zhouetal2014}].

Other dependent processes could be considered for $\mathcal{G}^{\mathcal
{X}}$, but a highly limiting requirement is that some aspect of the
location, for example, mean or median, can be fixed. There are few
examples where the process changes smoothly with covariates; one is the
multivariate beta process of \citet{Trippaetal2011}. Another approach
using Dirichlet process mixtures can be found in \citet{Reichetal2010},
but this latter approach would have to be extended to allow the means
or variances of the two mixture components to change with covariates.

\subsection{Posterior computation}\label{sec2.2}
The conditional likelihood for $(\bxi, \lambda_0, \be)$ is given by
%
\[
\mathcal{L}(\bxi, \lambda_0, \be)=\prod
_{i=1}^{n}\prod_{j=1}^{n_i}
\bigl[ \lambda_0(t_{ij}) \exp\bigl(\bw_{ij}'
\bxi+ e_i\bigr) \bigr] ^{\delta_{ij}} \exp\bigl\{-
\Lambda_0(t_{ij})\exp\bigl(\bw_{ij}'
\bxi+ e_i\bigr) \bigr\},
\]
where $\Lambda_0(t)=\int_0^t \lambda_0(s)\,ds$ is the cumulative hazard
function. The piecewise exponential model provides a flexible framework
to deal with the baseline hazard [see, e.g., \citet{WalkerMallick1997}].
We partition the time period $\mathbb{R}_+$ into $K$ prespecified
intervals, say, $I_k=(a_{k-1}, a_k], k=1, \ldots, K$, where $a_0=0$ and
$a_K=\max\{t_{ij}\}$. The baseline hazard is assumed to be constant
within each interval, that is,
\[
\lambda_0(t) = \sum_{k=1}^K
\lambda_k I\{t\in I_k \},
\]
where $\lambda_1, \ldots, \lambda_K$ are unknown hazard values and $I\{
A\}$ is the usual indicator function, that is, $1$ when $A$ is true,
$0$ otherwise. The prior hazard is specified by the hazard values $\{
\lambda_k\}_{k=1}^K$ and cut-point vector $\ba=(a_1, \ldots, a_K)$. If
the prior on the $\lambda_k$'s is taken to be independent gamma
distributions and $\{I_k\}_{k=1}^K$ is a reasonably fine mesh, the
gamma process [\citet{Kalbfleisch1978}] is approximated. To determine
the cut-point vector $\ba$, one can set $a_k$ to be the $\frac{k}{K}$th
quantile of the empirical distribution of the $t_{ij}$'s, or choose them
based on other considerations (see Section~\ref{SEERModelfit}). Some
authors have considered random cut-points [see, e.g., \citet
{SahuDey2004}]. Regardless, the resulting model implies a Poisson
likelihood [\citet{LairdOlivier1981}] as follows. Let $K(t)=\min\{k\dvtx
a_k \geq t\}$, $\Delta_{k}(t) = \min\{a_k, t\} - a_{k-1}$, and
$y_{ijk}=\delta_{ij}I\{k=K(t_{ij})\}$. Set $\bz_{ijk}=(\bolds{\iota
}_k', \bw_{ij}')'$ and $\bgamma=(\blambda', \bxi')'$, where $\bolds
{\iota}_k$ is a $K$-dimensional vector of zeros except the $k$th
element is $1$ and $\blambda=(\log(\lambda_1), \ldots, \log(\lambda
_K))'$. Then the likelihood for $(\bgamma,\be)$ becomes
%
\begin{eqnarray*}
\mathcal{L}(\bgamma, \be) &=& \prod
_{i=1}^{n}\prod_{j=1}^{n_i}
\bigl[ \exp\bigl\{ \log(\lambda_{K(t_{ij})}) + \bw_{ij}'
\xi+ e_i \bigr\} \bigr] ^{\delta_{ij}}
\\
&&\hspace*{29pt}{}\times \Biggl[ \prod
_{k=1}^{K(t_{ij})} e^{ -\exp \{ \log(\lambda_k) + \bw_{ij}'\xi+ e_i
\} \Delta_{k}(t_{ij}) } \Biggr]
\\
&=&\prod_{i=1}^{n}\prod
_{j=1}^{n_i} \prod_{k=1}^{K(t_{ij})}
\bigl[ \bigl( \exp\bigl\{ \bz_{ijk}'\bgamma+
e_i \bigr\} \bigr) ^{y_{ijk}} e^{ -\exp \{\bz_{ijk}'\bgamma+ e_i +
\log(\Delta_{k}(t_{ij})) \} } \bigr]
\\
& \propto&\prod_{i=1}^{n}\prod
_{j=1}^{n_i} \prod_{k=1}^{K(t_{ij})}
p(y_{ijk}|\bgamma, e_i),
\end{eqnarray*}
where\vspace*{1pt} $\mu_{ijk}=\exp \{ \bz_{ijk}'\bgamma+ e_i + \log(\Delta
_{k}(t_{ij})) \}$ and $p(y_{ijk}|\bgamma, e_i)$ is the probability mass
function for a Poisson distribution with mean $\mu_{ijk}$. For each
$i=1,\ldots,n$, let $N_i=\sum_{j=1}^{n_i}{K(t_{ij})}$, $\by
_i=(y_{ijk})$ be an\vspace*{1pt} $N_i \times1$ vector with subscript $ijk$ in
lexicographical order.

Thus, the proposed covariate-adjusted frailty PH model takes the
following hierarchical structure:
%
\begin{eqnarray*}
\by_i | \bgamma, e_i &\stackrel{\mathrm{ind}.} {\sim}&
\prod_{j=1}^{n_i} \prod
_{k=1}^{K(t_{ij})} p(y_{ijk}|\bgamma,
e_i),
\\
\bgamma&\sim& N_{K+p}(\bgamma_0,
\bS_0),
\\
e_i | G_{\bx_i} &\stackrel{\mathrm{ind}.} {\sim}&
G_{\bx_i},
\\
\mathcal{G}^{\mathcal{X}} \mid J,h,\theta, c,\rho&\sim&\operatorname{LDTFP}
\bigl(h,\Pi^{J,\theta},\mathcal{A}^{J,c,\rho}\bigr),
\\
\theta^{-2}&\sim&\Gamma(\tau_1,
\tau_2),\qquad c \sim\Gamma(a_c, b_c),
\end{eqnarray*}
which largely simplifies computations, where $N_p(\bmmm,\bS)$ refers to
a $p$-variate normal distribution with mean $\bmmm$ and covariance
matrix $\bS$. This forms the basis of an efficient Markov chain Monte
Carlo (MCMC) scheme for obtaining posterior inference, which can be
implemented using available software for generalized linear mixed
models. A full description of the MCMC algorithm is given in Appendix~B
of the supplementary material [\citet{Zhouetal2014}]. Sample R code
using the \texttt{LDTFPglmm} function available in \texttt{DPpackage}
[\citet{Jaraetal2011}] is provided in Appendix~C of the supplementary
material [\citet{Zhouetal2014}].

Time-dependent subject-specific covariates that are step-processes
[\citet{Hansonetal2009}] are naturally accommodated by including the
times where the covariate values change across all subjects into the
cut-point vector $\mathbf{a}$. All that is changed above is $\bz
_{ijk}=(\bolds{\iota}_k', \bw_{ijk}')'$, that is, $\bw_{ij}$ is
replaced with its time-varying analogue $\bw_{ijk}$. Similarly,
time-varying regression effects can be included by replacing $\bz
_{ijk}'\bgamma$ with $\bz_{ijk}'\bgamma_k$ in $\mu_{ijk}$, yielding
very general models. The proposed model implies exchangeable frailties
for each subgroup with a unique $\bx\in\mathcal{X}$. Time-dependent
cluster-specific covariates are therefore naturally included in the
model by simply allowing $\bx$ to change with time. For example, in the
SEER data set analyzed over a larger time window, for subjects living
in the $i$th county, one could include into $\bx_i$ the median house
income of that county at their particular diagnosis year. Furthermore,
the frailty distribution can itself evolve in time by simply including
time as a covariate in $\bx$, or a time-by-cluster covariate
interaction could also be entertained.

\section{Analysis of SEER county-level breast cancer data}\label{sec3}

\subsection{The Iowa SEER data}\label{sec3.1}

The SEER program of the National Cancer Institute (see \url{http://seer.cancer.gov/}) is an authoritative source of information on
cancer incidence and survival in the US, providing county-level cancer
data on an annual basis for particular states for public use. We fit
our proposed covariate-adjusted frailty Cox's PH model to a subset of
the Iowa SEER breast cancer survival data, which consists of a cohort
of 1073 women from the 99 counties of Iowa, who were diagnosed with
malignant breast cancer in 1995, with enrollment and follow-up
continued through the end of 1998. The observed survival time, from 1
to 48, was calculated as the number of months from diagnosis to either
death or the last follow-up. In our analysis, only deaths due to
metastasis of cancerous nodes in the breast were considered to be
events, while the deaths from other causes were censored at the time of
death. That is, we consider cause-specific survival models assuming
that all other deaths are independent of breast cancer. By the end of
1998, a total of 488 patients (45.5\%) had died of breast cancer, while
the remaining 585 patients were censored, either because they died of
other causes or survived until the last follow-up.

%
\begin{table}
\tabcolsep=0pt
\caption{Iowa SEER data: Summary statistics for follow-up times and
both individual- and county-level covariates}\label{tsummary}
\begin{tabular*}{\tablewidth}{@{\extracolsep{\fill}}@{}ld{2.3}d{3.3}d{3.3}@{}}
\hline
\textbf{Continuous variables} & \multicolumn{1}{c}{\textbf{Minimum}} & \multicolumn{1}{c}{\textbf{Median}} & \multicolumn{1}{c@{}}{\textbf{Maximum}} \\
\hline
Follow-up time in months & 1 & 19 & 47 \\
Age in years & 26 & 72 & 103 \\
RUCC & 2 & 7 & 9 \\
Income ($\times$1000) & 20.627 & 29.110 & 39.356
\\[9pt]
\textbf{Categorical variables} & \multicolumn{1}{c}{\textbf{Level}} & \multicolumn{1}{c}{\textbf{Count}} & \multicolumn{1}{c@{}}{\textbf{Proportion (\%)}} \\
\hline
Status & \multicolumn{1}{c}{Event} & \multicolumn{1}{c}{488} & 45.5 \\
& \multicolumn{1}{c}{Censored} & \multicolumn{1}{c}{585} & 54.5 \\[3pt]
Stage & \multicolumn{1}{c}{Local} & \multicolumn{1}{c}{510} & 47.5 \\
& \multicolumn{1}{c}{Regional} & \multicolumn{1}{c}{355} & 33.1 \\
& \multicolumn{1}{c}{Distant} & \multicolumn{1}{c}{208} & 19.4 \\
\hline
\end{tabular*}
\end{table}

\begin{figure}

\includegraphics{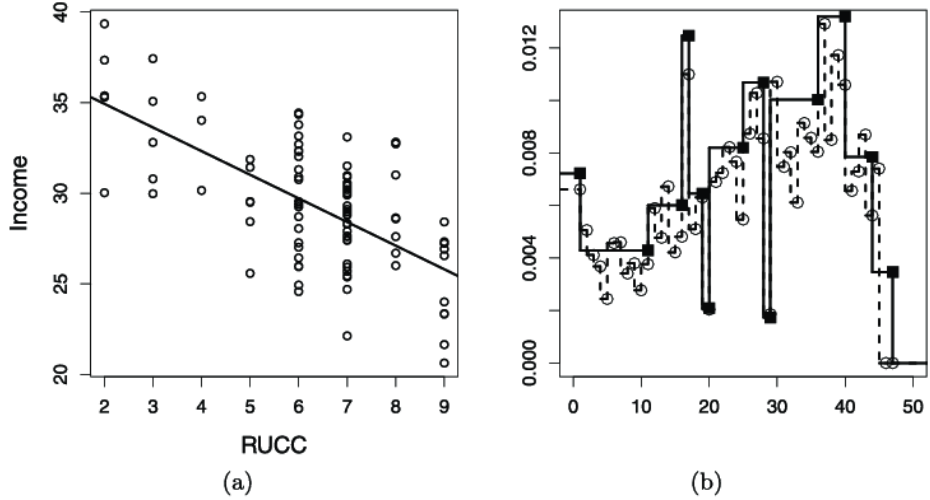}

\caption{Iowa SEER data:
panel \textup{(a)} shows the scatter plot and simple
linear regression line by regressing median household income on RUCC.
Panel \textup{(b)} shows the baseline hazards for {Model~1}. The dashed line
corresponds to Breslow's estimate of $\lambda_0(t)$ obtained by the GF
approach, where the circles represent the hazard values at each month;
the solid line is the fitted baseline hazard by our approach, where the
solid squares correspond to the cut-point values $\ba
=(1,11,16,17,19,20,25,28,29,36,40,44,47)$.}\label{LinearRegression}
\end{figure}

For each patient, the observed survival time and county of residence at
diagnosis are recorded. The data set also has individual-level
covariates including age at diagnosis and the stage of the breast
cancer: local (confined to the breast), regional (spread beyond the
breast tissue), or distant (metastasis). We create two dummy variables
for regional and distant, respectively, and treat the patients in the
local group as the baseline. Although several individual-level
covariates that affect breast cancer survival are not available (e.g.,
age at first child, age at menopause and breastfeeding), we are able to
obtain county-level covariates potentially associated with breast
cancer survival from census data, such as median household income
(small area estimates in 1993), poverty level (percentage of families
in poverty in 1990), education (percentage with Bachelor's degree or
higher in 1990) and rurality (Rural--Urban Continuum Codes in 1993). The
Economic Research Service Rural--Urban Continuum Codes (RUCC) vary from
1 to 9 (see
\href{http://www.ers.usda.gov/data-products/rural-urban-continuum-codes}{www.ers.usda.gov/data-products/rural-urban-continuum-codes}),
distinguishing metropolitan counties by the population size of their
metro area and nonmetropolitan counties by degree of urbanization and
adjacency to a metro area. Higher RUCC indicates a more rural county.
Other county-level covariates mentioned above are available at \url{http://data.iowadatacenter.org/browse/counties.html}. Since the
effects of education and poverty on the survival times are not
significant based on our initial model fitting by the proposed method,
we exclude them in the analysis presented below. Thus, we have
three-dimensional $\tilde{\bw}_{ij}$ and two-dimensional $\bx_{i}$.
Table~\ref{tsummary} presents several summary statistics for the data.
As shown in Figure~\ref{LinearRegression}, median household income and
RUCC are significantly, negatively correlated.

%

To get an initial feeling about the role that each county-level
covariate is playing, Table~\ref{tcontingency} provides the
distribution of each county-level covariate stratified by the
individual-level stage of disease. The gamma statistic (GK), originally
proposed by \citet{GoodmanKruskal1954}, is calculated to quantify the
association between each county-level covariate and the stage of
disease. The GK values range from $-$1 (100\% negative association)
to $1$ ($100 \%$ positive association), where the value $0$ indicates
no association. We see that women with a distant-stage at diagnosis are
much more likely than those with a local-stage to live in counties with
a high degree of urbanization ($\mathrm{GK}=-0.11$; $95\%$ CI:
from $-$0.20  to $-$0.01), while the association between stage and
income is not significant ($\mathrm{GK}=0.04$; $95\%$ CI: from
$-$0.06 to $0.13$). These associations roughly imply that women living
in urban counties may suffer poorer survival, assuming that women in
distant-stage are more likely to die than women in other stages.\vadjust{\goodbreak} Next,
we carefully examine both these individual-level and county-level
covariates in relation to breast cancer survival, fitting the
covariate-adjusted frailty proportional hazards model.

%
\begin{table}[b]
\tabcolsep=0pt
\caption{Iowa SEER data:
Distribution of each county-level covariate
stratified by individual-level stage. The pattern of numbers is Number
of women (\%). Goodman and Kruskal's gamma statistics (95\% confidence
intervals) are $-$0.11 $(-0.20, -0.01)$ and $0.04$ $(-0.06, 0.13)$ for
RUCC and Income,~respectively}\label{tcontingency}
\begin{tabular*}{\tablewidth}{@{\extracolsep{\fill}}@{}lcccc@{}}
\hline
& & \multicolumn{3}{c@{}}{\textbf{Stage}} \\[-4pt]
& & \multicolumn{3}{c@{}}{\hrulefill}\\
& \multicolumn{1}{c}{\textbf{All women}}
& \multicolumn{1}{c}{\textbf{Local}}
& \multicolumn{1}{c}{\textbf{Regional}}
& \multicolumn{1}{c@{}}{\textbf{Distant}}
\\
\multicolumn{1}{@{}l}{\textbf{Covariates}}
& \multicolumn{1}{c}{$\bolds{N=1073}$}
& \multicolumn{1}{c}{$\bolds{N=510}$}
& \multicolumn{1}{c}{$\bolds{N=355}$}
& \multicolumn{1}{c}{$\bolds{N=208}$}\\
\hline
{RUCC} & & & & \\
1--3 & 314 (29.3)& 131 (25.7) & \phantom{0}99 (27.9) & \phantom{0}84 (40.4) \\
4--7 & 666 (62.1)& 342 (67.1) & 221 (62.3) & 103 (49.5) \\
8--9 & 93 (8.6) & 37 (7.2) & 35 (9.8) & \phantom{0}21 (10.1)
\\[3pt]
{Income} ($\times$1000) & & & & \\
20--27 & 163 (15.2)& \phantom{0}79 (15.5) & \phantom{0}51 (14.4) & \phantom{0}33 (15.9) \\
27--34 & 651 (60.7)& 312 (61.2) & 223 (62.8) & 116 (55.8) \\
$>$34 & 259 (24.1)& 119 (23.3) & \phantom{0}81 (22.8) & \phantom{0}59 (28.3) \\
\hline
\end{tabular*}
\end{table}

\subsection{Models and model comparison}\label{sec3.2} \label{SEERModelfit}

We fitted the proposed covariate-adjusted frailty PH model for the Iowa
SEER data with different county-level covariates, including models with
RUCC only ({Model~1}), with median household income only ({Model~2})
and with both ({Model~3}). To see how the piecewise assumption of
baseline hazard affects the predictive ability of models, we considered
three specifications of cut-point vector $\ba$ as follows:
\begin{longlist}[\textit{Case} III]
\item[\textit{Case} I.] $\ba=(1,11,16,17,19,20,25,28,29,36,40,44,47)$, which
was determined by visually examining Breslow's estimate of $\lambda
_0(t)$ using the GF approach, which is given in panel~(b) of Figure~\ref{LinearRegression}.

\item[\textit{Case} II.] $\ba=(3,7,12,16,19,24,29,34,41,47)$, where $a_k$ is
the $\frac{k}{10}$th quantile of the empirical distribution of observed
survival times.

\item[\textit{Case} III.] $\ba=(47)$, which yields an exponential baseline hazard.
\end{longlist}

In all cases, we set $J=4$. We fitted all the models using the
corresponding variants of the algorithm described in Appendix~B of the
supplementary material [\citet{Zhouetal2014}] and similar prior
specifications suggested in the simulation study. The Markov chain
mixed reasonably well despite the high dimension of our models. For
each version of our model and case, we ran a single Markov chain of
1,020,000. A total number of 20,000 were discarded as burn-in
period and 10,000 samples were retained for posterior inference.
Moreover, we also considered another {case~II} with $13$ cut-points and
cut-point specifications based on the event time quantiles from the
Kaplan--Meier curve in Appendix~E of the supplementary material [\citet
{Zhouetal2014}]. The results show that carefully choosing the
cut-points is more important than simply increasing the number of cut-points.

For the sake of comparison, we further fitted the exchangeable MPT
frailty Cox model and the Bayesian exchangeable Gaussian frailty Cox
model. We compare the models using the log pseudo marginal likelihood
(LPML) developed by \citet{GeisserEddy1979} and the deviance information
criterion (DIC) proposed by \citet{Spiegelhalteretal2002}. In the
context of the frailty Cox model, the LPML for model $M$ is defined as
$\LPML= \sum_{i=1}^n \sum_{j=1}^{n_i} \log(\CPO_{ij})$, where $\CPO
_{ij}$, the $ij$th conditional predictive ordinate, is given by
$[\lambda(t_{ij} )^{\delta_{ij}} e^{-\Lambda(t_{ij})} | \calD_{(ij)}]$
with $\calD_{(ij)}$ denoting the remaining data after excluding the
$ij$th data point $\calD_{ij}$. One can use the simple method suggested
by \citet{GelfandDey1994} to estimate the CPO statistics from MCMC
output. A larger value of LPML indicates the corresponding model has
better predictive ability. Furthermore, \citet{GeisserEddy1979}
discussed the exponentiated difference in LPML values from two models
to obtain what they termed as a pseudo Bayes factor (PBF). The PBF is a
surrogate for the more traditional Bayes factor and can be interpreted
similarly, but is more analytically tractable, much less sensitive to
prior assumptions, and does not suffer from Lindley's paradox. Set
$\bOmega=(\be, \bgamma, \bbeta, \theta)$ as the entire collection of
model parameters. The DIC for model $M$ is defined as $\DIC=\bar{D}+p_D
=E_{\bOmega|\calD} \{D(\bOmega) \} + p_D$, where $D(\bOmega)=-2\log
\calL(\bgamma, \be)$ which is referred to as the deviance function, and
$p_D=\bar{D}-D (E_{\bOmega|\calD} \{\bOmega\} )$ which is a measure of
model complexity. Note that the DIC is also readily computed from MCMC output.

\subsection{Results}\label{sec3.3}

Table~\ref{tLPML} shows the DIC and LPML for all models under
consideration. All models under {case~I} provide significantly better
prediction as measured by both DIC and LPML, with differences in the
range of 20--55 for DIC and 10--25 for LPML, which indicates that the
determination of the cut-point vector for the baseline hazard plays an
important role on model prediction and fit. Comparing the frailty
specifications in {Model~1} across all cases, the DIC and LPML show the
same trend for goodness of fit, with the proposed model based on the
LDTFP frailty model outperforming both the MPT and Gaussian models,
although the differences are only in the range of 1--4. Comparing
between {Model~2} and {Model 3}, the proposed model is always preferred
in terms of LPML, while the MPT model is slightly better than others in
term of DIC under {Model~2}. Comparing all the proposed models across
{Model~1}--{Model 3}, the results indicate that {Model~1} always
performs best. Overall, allowing the frailty distribution to change
with county-level covariates (especially RUCC) does improve model
prediction according to LPML. In what follows, we present the results
under {case~I}.

%
\begin{table}
\tabcolsep=0pt
\caption{Iowa SEER data:
Deviance information criteria (DIC) and log
of the pseudo marginal likelihood (LPML) for models under consideration}\label{tLPML}
\begin{tabular*}{\tablewidth}{@{\extracolsep{\fill}}@{}lccccccc@{}}
\hline
& & \multicolumn{2}{c}{\textbf{Case I}} & \multicolumn{2}{c}{\textbf{Case II}} & \multicolumn{2}{c@{}}{\textbf{Case III}}\\[-4pt]
& & \multicolumn{2}{c}{\hrulefill} & \multicolumn{2}{c}{\hrulefill} & \multicolumn{2}{c@{}}{\hrulefill}\\
\textbf{Model} & \textbf{Frailty} & \textbf{DIC} & \textbf{LPML} & \textbf{DIC} & \textbf{LPML} & \textbf{DIC} & \textbf{LPML}\\
\hline
1 & LDTFP & 4436 & $-$2222 & 4463 & $-$2234 & 4495 & $-$2247 \\
& MPT & 4441 & $-$2225 & 4463 & $-$2235 & 4496 & $-$2248 \\
& Gaussian & 4444 & $-$2225 & 4467 & $-$2236 & 4497 & $-$2248
\\[3pt]
2 & LDTFP & 4441 & $-$2224 & 4465 & $-$2235 & 4498 & $-$2249 \\
& MPT & 4440 & $-$2225 & 4462 & $-$2236 & 4497 & $-$2248 \\
& Gaussian & 4443 & $-$2225 & 4465 & $-$2235 & 4498 & $-$2249
\\[3pt]
3 & LDTFP & 4438 & $-$2223 & 4464 & $-$2235 & 4496 & $-$2248 \\
& MPT & 4441 & $-$2225 & 4464 & $-$2235 & 4498 & $-$2249 \\
& Gaussian & 4445 & $-$2226 & 4467 & $-$2236 & 4498 & $-$2248 \\
\hline
\end{tabular*}
\end{table}

%
\begin{table}
\tabcolsep=0pt
\caption{Iowa SEER data: Posterior medians (95\% credible intervals)
of fixed effects $\bxi$ from various models}\label{tcoefficients}
\begin{tabular*}{\tablewidth}{@{\extracolsep{\fill}}@{}lccccc@{}}
\hline
\textbf{Predictor} & \textbf{Model~1} & \textbf{Model~2} & \textbf{Model 3} & \textbf{CAR} & \textbf{Cox}\\
\hline
$\xi_1$ (Age) & 0.019 & 0.020 & 0.020 & 0.018 & 0.019 \\
& (0.013, 0.025) & (0.014, 0.026) & (0.014, 0.026) & (0.012, 0.025) &(0.013, 0.025)
\\[3pt]
$\xi_2$ (Regional) & 0.27 & 0.27 & 0.27 & 0.22 & 0.30 \\
& (0.03, 0.49) & (0.03, 0.47) & (0.05, 0.50) & (0.01, 0.49) & (0.08, 0.52)
\\[3pt]
$\xi_3$ (Distant) & 1.64 & 1.67 & 1.65 & 1.65 & 1.64 \\
& (1.43, 1.88) & (1.43, 1.89) & (1.43, 1.89) & (1.40, 1.93) & (1.42, 1.87)
\\[3pt]
$\xi_{x_1}$ (RUCC) & $-$0.105 & & $-$0.082 & & \\
&($-$0.185, $-$0.041) & & ($-$0.179, 0.011) & &
\\[3pt]
$\xi_{x_2}$ (Income) & & 0.042 & 0.011 & & \\
& & (0.003, 0.084) & ($-$0.040, 0.066) & & \\
\hline
\end{tabular*}
\end{table}

Table~\ref{tcoefficients} presents posterior medians and equal-tailed
$95\%$ credible intervals (CI) for main effects (components of $\bxi$)
under {Model~1}--{Model 3}, with covariate-adjusted frailties, and
compares the individual-level covariate effects, that is, ($\xi_1, \xi
_2,\xi_3$), to those obtained by \citet{Zhaoetal2009}, under the
standard nonfrailty Cox model and the Cox frailty model that has a MPT
prior for the baseline survival, centered at the log-logistic family,
and with conditionally autoregressive (CAR) county-level spatial
frailties. The best fitting Cox model reported by \citet{Zhaoetal2009}
has an LPML of $-$2226. Therefore,\vspace*{1pt} the pseudo Bayes factor for the
proposed model versus the CAR model is $e^{2226\mbox{--}2222} \approx55$,
implying that the proposed model predicts about 55 times better than
the model with CAR frailties. In addition, the proposed model offers a
unique interpretation. The posterior medians and 95\% CIs for all
individual-level effects change little across the different versions of
the proposed model, indicating that the Cox regression estimates are
reasonably stable for these data, except for the estimate of ``Regional
stage,'' for which the CAR model 95\% CI is much wider than those under
the considered versions of the proposed model. This may be partly due
to the benefit of including county-level covariates. The best model
according to LPML, {Model~1}, indicates that all the individual-level
effects are significant at the 0.05 level. Higher age at diagnosis
increases the hazard within each county. For instance, women are about
$e^{0.019\times20}\approx1.46$ times more likely to die from breast
cancer than those twenty years younger who have the same disease stage
and live in the same county. Compared with women having local stage of
disease, women of the same age and living in the same county are
$e^{0.27}\approx1.31$ times more likely to die if their cancer is
detected at the regional stage, and $e^{1.64} \approx5.16$ times more
likely to die if detected at the distant stage. We additionally present
the fixed effects under the marginal PH model (i.e., using the R
function \texttt{coxph} with option \texttt{cluster}) across {Model~1}--{Model 3} in Appendix~E of the supplementary material [\citet
{Zhouetal2014}]. Note that the coefficient estimates under the marginal
PH model have population-averaged interpretations and cannot be
directly compared with those fitted from the proposed frailty PH model
due to different model structures. 

\begin{figure}

\includegraphics{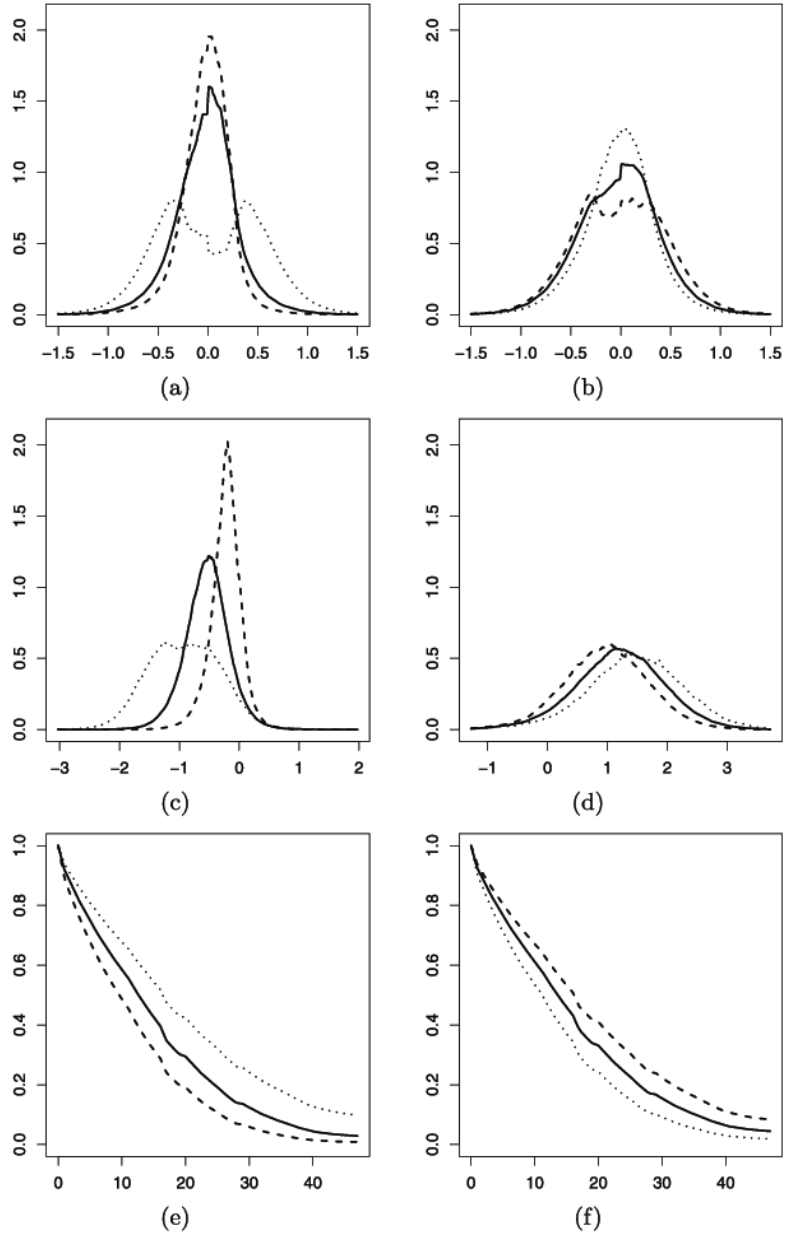}

\caption{Iowa SEER data: Fitted predictive frailty densities [panels
\textup{(a)} and \textup{(b)}], frailty densities with location shifts [panels \textup{(c)} and
\textup{(d)}] and survival curves [panels \textup{(e)} and \textup{(f)}] for women with mean entry
age 68.8 years and distant stage of disease from different county
covariate levels under Model~1 [panels \textup{(a)}, \textup{(c)} and \textup{(e)}] and Model~2
[panels \textup{(b)}, \textup{(d)} and \textup{(f)}]. In panels \textup{(a)}, \textup{(c)} and \textup{(e)}, the results for
$\mathrm{RUCC}=2$, $5$ and $9$ are displayed as dashed, continuous and dotted
lines, respectively. In panels \textup{(b)}, \textup{(d)} and \textup{(f)}, the results for
$\mathrm{Income}=23.354$, $29.176$ and $35.301$ are displayed as dashed,
continuous and dotted lines, respectively.}
\label{model1&2}
\end{figure}

\begin{figure}

\includegraphics{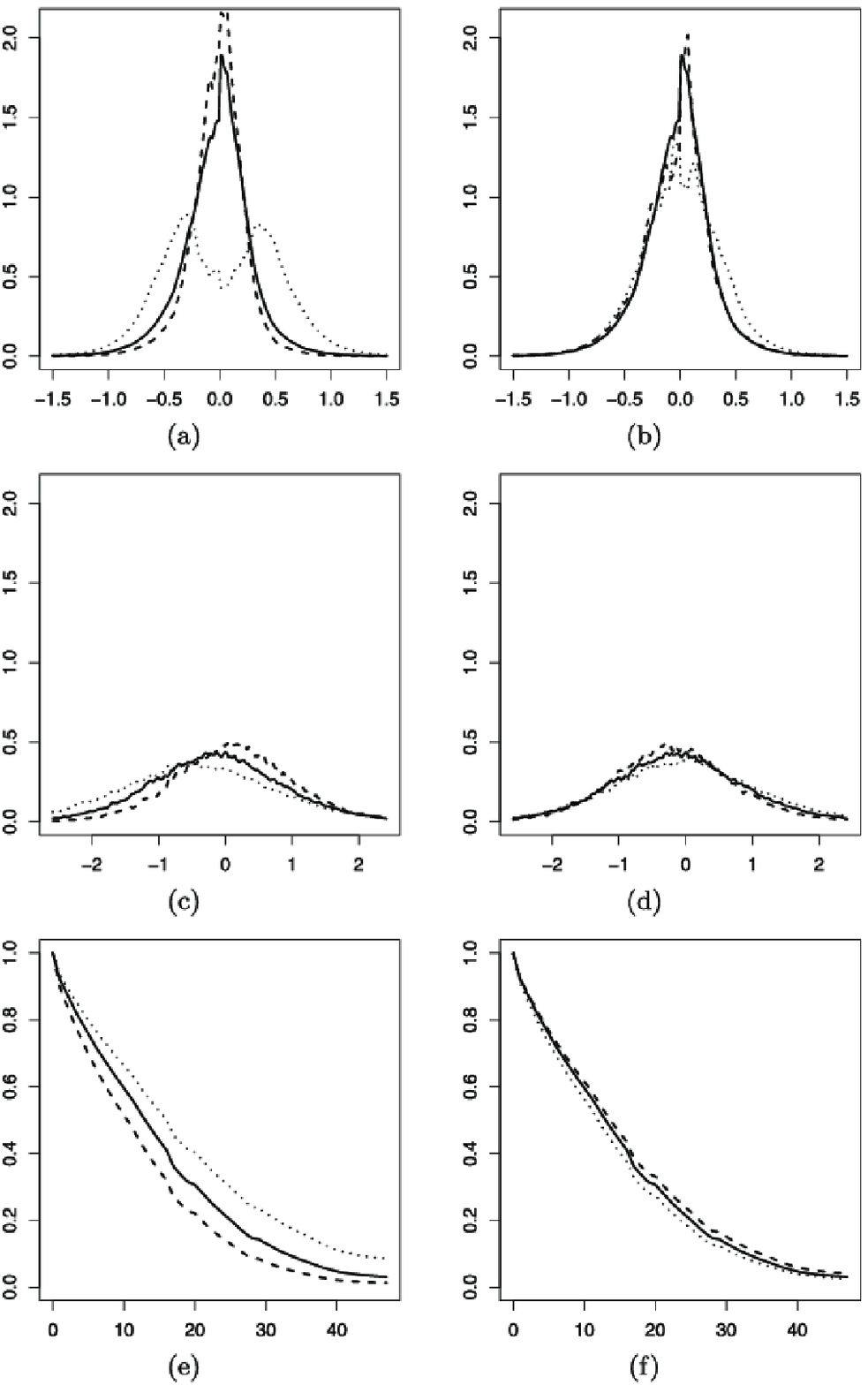}

\caption{Iowa SEER data: Fitted predictive frailty densities [panels
\textup{(a)} and \textup{(b)}], frailty densities with location shifts [panels \textup{(c)} and
\textup{(d)}] and survival curves [panels \textup{(e)} and \textup{(f)}] for women with mean entry
age 68.8 years and distant stage of disease from different county
covariate levels under Model~3. In panels \textup{(a)}, \textup{(c)} and \textup{(e)}, the results
for $\mathrm{RUCC}=2$, $5$ and $9$ are displayed as dashed, continuous and
dotted lines, respectively. In panels \textup{(b)}, \textup{(d)} and \textup{(f)}, the results for
$\mathrm{Income}=23.354$, $29.176$ and $35.301$ are displayed as dashed,
continuous and dotted lines, respectively.}
\label{model3}
\end{figure}

Regarding the effect of county-level covariates, living in a higher
median household income or urban counties is associated with poorer
survival after a breast cancer diagnosis. For example, the results
under {Model~1} indicate that after controlling for individual
covariates and county,\vspace*{1pt} the hazard rate of women living in urban
counties (with $\mathrm{RUCC}=2$) will be $e^{0.105\times7}
\approx2$ times larger than that of women in rural counties (with
$\mathrm{RUCC}=9$). The results under {Model~2} imply that after
controlling for individual covariates and frailties, women have about a
$1.7$ times larger hazard rate if they live in median household income
counties of \$35,301 compared with median household income of \$23,354 (see also Figure~\ref{model1&2}). Under {Model 3}, the results
indicate that when both the county-level covariates are included
simultaneously, their independent effects are attenuated, partly due to
the multicollinearity between them (see the middle two plots in
Figure~\ref{model3}).

%

%

We obtain the fitted predictive frailty densities for both $\be_i$
(median-zero) and $\be_i+\bx_i'\bxi_x$ (full distribution) and survival
curves for women with mean entry age $68.8$ years and distant stage of
disease who live in the counties with different levels of median
household income or RUCC, under the different versions of the proposed
model. The three levels are chosen from the 5\%, 50\% and 95\%
quantiles of each covariate value. The results are reported in
Figures~\ref{model1&2} and~\ref{model3}. Under our best fitting, {Model~1} (see left three plots in Figure~\ref{model1&2}), the results
indicate that higher values of RUCC increase the frailty variance and
suggest a non-Gaussian shape (upper); we also see overall higher
frailty after mixing over the location shift $\bx_i'\bxi_x$ (middle)
and so poorer survival (lower) in urban counties. Increasing
heterogeneity as ruralness increases under {Model~1} translates
into increasing association among those living in more rural counties
versus urban. In Appendix~E of the supplementary material [\citet
{Zhouetal2014}], Kendall's tau is computed and plotted as a function of
RUCC for individuals with mean entry age $68.8$ years and distant
stage. Kendall's tau increases \emph{by a factor of three} as RUCC goes
from 2 to 9. Note that under a traditional gamma frailty model the
association is static.

Under {Model~2}, the frailty densities only slightly change compared
with {Model~1}, but we do see poorer survival in counties with higher
median household income. Figure~\ref{model3} demonstrates that after
adjusting individual covariates and median household income (right
three plots), there is little effect of RUCC on either predictive
frailty densities or survival curves; while after adjusting for RUCC
(left three plots), the effect of median household income is almost
negligible. In Appendix~E of the supplementary material [\citet
{Zhouetal2014}], the survival curves are also compared with those
obtained under the marginal PH model. Overall, the marginal PH model
under-predicts survival time up to about $1$ month, for example, it
gives estimates of median survival a month less, compared with our
proposed model for patients with mean entry age $68.8$ years and
distant stage of disease who live in the same county. This may be
partly due to the fact that the marginal PH model averages over the
changing behavior of the frailty distribution over the ruralness measure.

It is widely known that access to quality care and screening for breast
cancer is more readily available to those with greater financial means
and/or those living in urban areas. Therefore, our findings of
increased survival for poorer and more rural counties for this cohort
are initially puzzling. However, hormone replacement therapy (HRT)
increased about 150\% in the 1990s [\citet{WysowskiGovernale2005}],
after several observational studies linked HRT to prevention of
osteoporosis and protection from heart disease. However, this
increasing use of HRT abated suddenly in 2002, when the Women's Health
Initiative clinical trial linked HRT to aggressively invasive breast
cancer [\citet{Rossouwetal2002}]. In fact, overall breast cancer
incidence rates peaked in 1999. Between 2001 and 2004 overall invasive
breast cancer incidence declined, but fell much more drastically among
women living in urban versus rural counties, and among women living in
low-poverty versus high-poverty counties. \citet{Hausaueretal2009}
attribute this discrepancy to greater use of postmenopausal
estrogen/progestin hormone replacement therapy among more affluent
women and/or women living in urban counties up until 2002, when the
Women's Health Initiative trial was stopped prematurely on May 31,
2002, according to \citet{Rossouwetal2002}, ``\ldots\emph{because the
test statistic for invasive breast cancer exceeded the stopping
boundary for this adverse effect and the global index statistics
supported risks exceeding benefits.}'' It is plausible that increased
risk (i.e., stochastically larger frailties) in more affluent and more
urban counties has to do with a larger proportion of women being
prescribed HRT in the late 1980s and 1990s. Further exploratory
analyses on other cohorts of SEER Iowan breast cancer data (1975--1979,
1980--1984, 1985--1989 and 1990--1994) show a reversal of the effects
of income and ruralness, agreeing with intuition. Paralleling our
study, \citet{Kriegeretal2010} used county-level census data on income
and found rising and falling breast cancer incidence rates for the SEER
data over the range 1992--2005 for caucasian women living in
high-income counties, which ``\emph{mirrored the social patterning of
hormone therapy use.}''

In a longer follow-up study of the Women's Health Initiative trial,
\citet{Chlebowskietal2010} found that those on estrogen plus progestin
compared to placebo had about 25\% higher incidence of invasive breast
cancer. Among those diagnosed with breast cancer, the two treatment
arms had similar histology, but the estrogen plus progestin group were
78\% more likely to have cancers that had spread to lymph nodes than
placebo, and the estrogen plus progestin group were about twice as
likely to die from breast cancer versus placebo. It would appear that
hormone replacement therapy fortified the virulence of breast
cancer, significantly increasing both incidence and mortality. This
same study showed an impressive 7\% one-year drop in incidence right
after the Women's Health Initiative study was prematurely stopped and
the medical community warned of a possible link between hormone
replacement therapy and breast cancer.

\section{Simulation studies}\label{sec4}
We performed a simulation study to assess the performance of the
proposed model. The simulated data are also used to compare the
proposed approach with existing models. Specifically, we consider the
GF approach described in Section~\ref{sec2.1} and the positive stable frailty
Cox model proposed by \citet{Liuetal2011}.
Under this latter model, the shape parameter is allowed to depend on
cluster-level covariates. In terms of our notation, they assumed that
the conditional hazard function of $T_{ij}$ is
%
\begin{equation}
\label{PSF} \lambda(t|\tilde{\bw}_{ij},\bx_i,
e_i) = \lambda_{0i}(t)\exp\bigl(\tilde{
\bw}_{ij}'\tilde{\bxi}_i + e_i
\bigr),
\end{equation}
where the baseline hazard functions $\lambda_{0i}(t)$ and regression
parameters $\tilde{\bxi}_i$ are cluster-specific, and $\exp(e_i)$
follows a positive stable distribution with shape parameter $\alpha_i
\in(0,1)$, relying on the cluster-level covariates vector $\bx_i$
through a logit link function, denoted by $PS(\alpha_i)$. They did not
deal with this model directly, but rather derived the marginal model
%
\begin{equation}
\label{PSFmarginal} \lambda(t|\tilde{\bw}_{ij},\bx_i) =
h_{0}(t)\exp\bigl(\tilde{\bw}_{ij}'\boeta\bigr)
\end{equation}
by imposing the restrictions $\boeta=\alpha_i\tilde{\bxi}_i$, $H_0(t)=\{
\Lambda_{0i}(t)\}^{\alpha_i}$, where $H_0(t) = \int_0^t h_0(s)\,ds$ and
$\Lambda_{0i}=\int_0^t \lambda_{0i}(s)\,ds$. In other words, they
essentially fitted the above marginal Cox model by maximizing the
pseudo partial likelihood under the working independence assumption
[\citet{Weietal1989}], and then utilized the imposed constraints to
estimate the parameters in the frailty model. Although they considered
a more flexible conditional Cox model, they made many assumptions to
get the marginal model, some of which are difficult to check in
practice. Moreover, they faced a nonidentifiability problem when a
cluster-level covariate was included in the conditional Cox model, so
cluster-level covariates had to be excluded from the marginal model as
well, leading to potentially poorer prediction of the marginal survival
function. Their method, referred to below as PSF, will be compared with
our approach focusing on the prediction of survival functions in the
simulation studies. A comparison of the two methods for the fixed
effect estimates cannot be conducted, since they have different model
structures. We conducted the simulation study in \verb!R!. The GF and
PSF approaches were implemented by using the function \texttt{coxme}
and \texttt{coxph} (with the option \texttt{cluster}), respectively,
included in the R packages \pkg{coxme} and \pkg{survival}.

\subsection{Simulation settings}\label{sec4.1}
Two scenarios for the frailty distributions were considered. In the
first case, referred to as {Scenario I}, a covariate-dependent family
of distributions is considered, where the density shape evolves from
one mode to two as the cluster-specific covariate $x$ increases its
value; this mirrors the effect of RUCC in panel \textup{(a)} of Figure~{\ref
{model3}} for {Model~1}. In the second case, referred to as {Scenario
II}, a covariate-dependent positive stable distribution is considered.
The specific distributional forms for each setting were the following:
\begin{longlist}[\textit{Scenario} II.]
\item[\textit{Scenario} I.] $e_i|x_i \stackrel{\mathrm{ind}.}{\sim}
0.5N(-e^{0.4x_i},1) + 0.5N(e^{0.4x_i},1)$, $x_i \stackrel{\mathrm{i.i.d.}}{\sim}
U(-3,3)$.

\item[\textit{Scenario} II.] $\exp(e_i)|x_i \stackrel{\mathrm{ind}.}{\sim} PS(\alpha
_i)$, $\alpha_i=1/ (1+e^{-0.5\mbox{--}0.5x_i} )$, $x_i \stackrel{\mathrm{i.i.d.}}{\sim} U(0,2)$.
\end{longlist}
Note that the first setting is not a particular case of the proposed
model; the second setting, chosen from the simulation study of \citet
{Liuetal2011}, is included to evaluate the behavior of the proposed
approach when the PSF model is correct.

Given the frailties, the data under {Scenario I} were simulated from
the conditional PH model (\ref{frailtyPH}) with $\lambda_0(t)=1$, $\bw
_{ij}=({w}_{1ij}, {w}_{2ij}, x_i)'$ and $\bxi=(\xi_1,\xi_2,\xi
_x)'=(1.0,0.5,1.0)'$; the data under {Scenario II} were simulated from
the PSF model (\ref{PSF}) with $\tilde{\bw}_{ij}=(w_{1ij}, w_{2ij})'$,
$\boeta=(1,0.5)'$ and $H_0(t)=t$. For each simulation scenario, $200$
replicates of the data set were generated by assuming the following:
$w_{1ij} \stackrel{\mathrm{i.i.d.}}{\sim} N(0,1)$ and $w_{2ij} \stackrel
{\mathrm{i.i.d.}}{\sim}\mbox{Bernoulli}(0.5)$, $i=1,\ldots,100$, $j=1,\ldots
,10$. In each case, a noninformative censoring scheme was considered,
where the censoring times were simulated from an $U(0.25,4)$
distribution, so that the censoring rate is approximately $35\%$ under
{Scenario I} and $25\%$ under {Scenario II}.

For each data set, the GF approach was employed, yielding point
estimates of~$\bxi$, $\operatorname{var}(e_i)$ and $e_i$, which we
denote by $\hat{\bxi}^{(0)}$, $\hat{\theta}^{2(0)}$ and $\hat
{e}_i^{(0)}$, respectively. Based on these point estimates, the
predictive survival function was calculated as follows:
%
\begin{equation}
\label{SurvivalGF} \hat{S}_{\mathrm{GF}}(t|\bw) = n^{-1} \sum
_{i=1}^{n} \exp\bigl\{ -\hat{\Lambda}_0^{(0)}(t)
\exp\bigl\{\bw'\hat{\bxi}^{(0)} + \hat{e}_i^{(0)}
\bigr\} \bigr\},
\end{equation}
where $\hat{\Lambda}_0^{(0)}(t)$, depending on $\hat{e}_i^{(0)}$'s,
denotes Breslow's estimator of $\Lambda_0(t)$ [see, e.g., \citet
{Therneauetal2003}, Section~2]. We then fitted the proposed model, by
considering $J=4$, $K=10$, $\tau_1=1.001$, $\tau_2=1.001\hat{\theta
}^{2(0)}$, $a_c=1$, $b_c=1$, $\bgamma_0=\mathbf{0}_{13}$ and $\bS
_0=10^3\times\bI_{13}$. For each data set a single Markov chain of
length 55,000 was obtained by using the algorithm described in
Appendix~B of the supplementary material [\citet{Zhouetal2014}]. A
burn-in period of 5000 scans was considered, and 5000 samples
were retained for posterior inferences. The posterior mean of the
corresponding parameters are denoted by $\hat{\bxi}$, $\hat{\theta}^2$,
$\hat{g}(e|\bx)$ and $\hat{S}(t|\bw)$. Finally, the PSF approach was
considered but including the cluster-level covariates in the linear
predictor, and the associated predictive survival function, based on
Breslow's estimator of the underlying baseline hazard function, was
obtained and is denoted by $\hat{S}_{\mathrm{PSF}}(t|\bw)$.

The competing approaches were compared regarding the estimation of the
regression coefficients and also compared by computing the weighted
integrated squared error (ISE) for the estimated survival
distributions, given by
\[
\displaystyle \int_0^{\infty} \bigl\{ \hat{S}_m(t|\bw)-S(t|\bw) \bigr\}^2 f_T(t|\bw)\,dt,
\]
where $\hat{S}_m(t|\bw)$, $S(t|\bw)$ and $f_T(t|\bw)$ are the estimated
survival function, the true survival function and density function,
respectively, for a subject with covariate vector $\bw$.

%
\begin{table}[b]
\tabcolsep=0pt
\caption{Simulation data---Scenario \textup{I}: True value, bias of the point
estimator, mean (across Monte Carlo simulations) of the posterior
standard deviations/standard errors (MEAN-SD), standard deviation
(across Monte Carlo simulations) of the point estimator (SD-MEAN) and
Monte Carlo coverage probability for the $95\%$ credible
interval/confidence interval (CP) for the regression parameters. The
results are presented under the proposed model and under the GF approach}\label{tone}
\begin{tabular*}{\tablewidth}{@{\extracolsep{\fill}}@{}lccccccccc@{}}
\hline
& & \multicolumn{4}{c}{\textbf{Proposed model}} & \multicolumn{4}{c@{}}{\textbf{GF model}}\\[-4pt]
\multirow{2}{26pt}{\textbf{Para\-meters}}& & \multicolumn{4}{c}{\hrulefill} & \multicolumn{4}{c@{}}{\hrulefill}\\
 & \textbf{True} & \textbf{BIAS} & \textbf{MEAN-SD} & \textbf{SD-MEAN} & \textbf{CP} & \textbf{BIAS} & \textbf{MEAN-SD} & \textbf{SD-MEAN} & \textbf{CP} \\
\hline
$\xi_1$ & 1.0 & \phantom{$-$}0.011 & 0.052& 0.054& 0.930 &  $-$0.011& 0.051& 0.059& 0.910 \\
$\xi_2$ & 0.5 & \phantom{$-$}0.008 & 0.088& 0.090& 0.945 &  $-$0.003& 0.088& 0.091& 0.950 \\
$\xi_x$ & 1.0 & $-$0.009& 0.141& 0.126& 0.965 &  $-$0.052& 0.083& 0.142& 0.775 \\
\hline
\end{tabular*}
\end{table}

\subsection{Simulation results}\label{sec4.2}
The results for the regression coefficients using the proposed model
and the GF approach under {Scenario I} are given in Table~\ref{tone},
where the bias of the corresponding point estimators, the Monte Carlo
mean of the posterior standard deviation/standard error (MEAN-SD), the
Monte Carlo standard deviation of the point estimates (SD-MEAN) and the
Monte Carlo coverage probability (CP) of the $95\%$ credible
interval/confidence intervals are presented. The results suggest that
the posterior means of $\bxi$ are almost unbiased estimators and that
the observed bias for $\xi_x$ under the proposed approach is much
smaller than the corresponding value obtained under the GF approach.
Moreover, under the proposed model, the MEAN-SD and the SD-MEAN values
are in fairly close agreement, indicating that the posterior standard
deviation is an unbiased estimator of the frequentist standard error.
Finally, the CPs are all around the nominal $95\%$. The same does not
hold for GF, which substantially underestimates the standard error for
$\xi_x$, leading to low coverage probabilities. 

\begin{figure}

\includegraphics{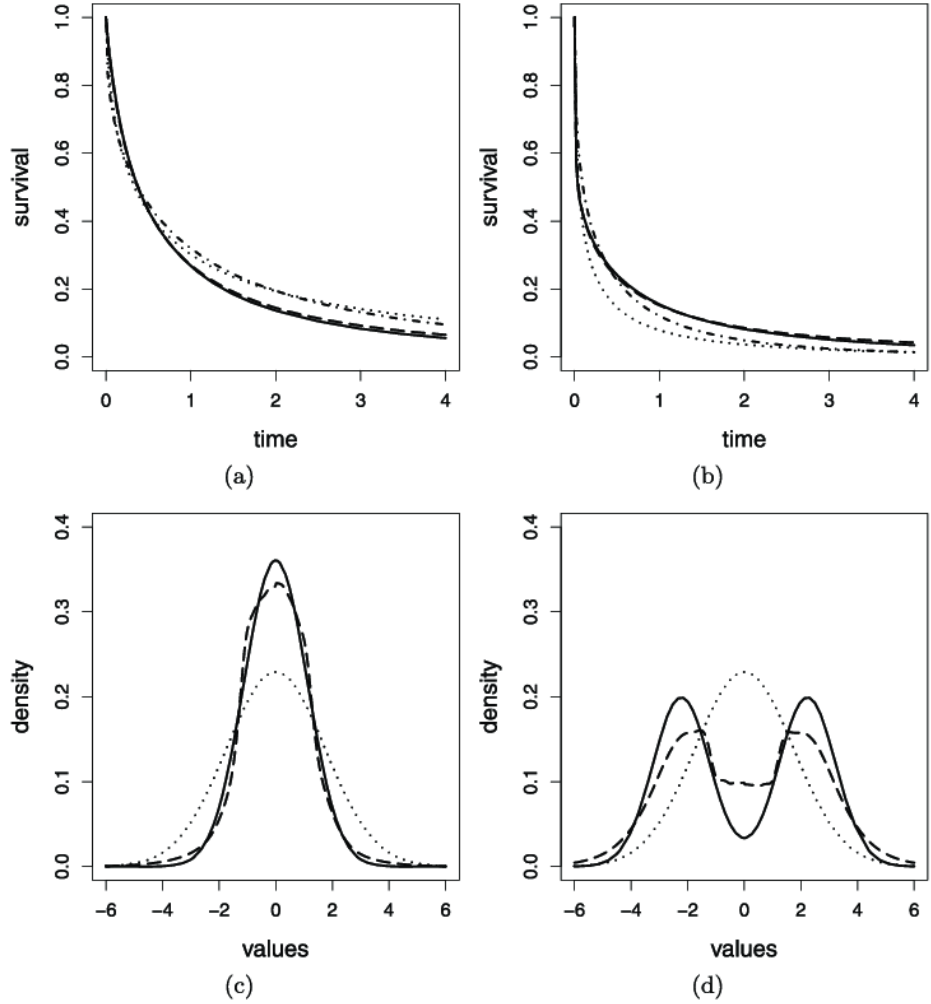}

\caption{Simulated data---Scenario \textup{I}: Mean, across simulations, of the
posterior mean of the survival and frailty density functions under the
proposed model. Panels \textup{(a)} and \textup{(b)} show the results for the survival
functions. Panels \textup{(c)} and \textup{(d)} show the results for the frailty
densities. Panels \textup{(a)} and \textup{(c)} show the results for covariate values
$(2,1,-2)$. Panels \textup{(b)} and \textup{(d)} show the results for covariate values
$(0,1,~2)$. The true curves are represented by continuous lines. The
results under the proposed model are represented by dashed lines. The
results under the exchangeable Gaussian frailty model are represented
by dotted lines. In panels \textup{(a)} and \textup{(b)} the results obtained under the
PSF approach are represented by dot-dashed lines.}
\label{case1}
\end{figure}

\begin{figure}

\includegraphics{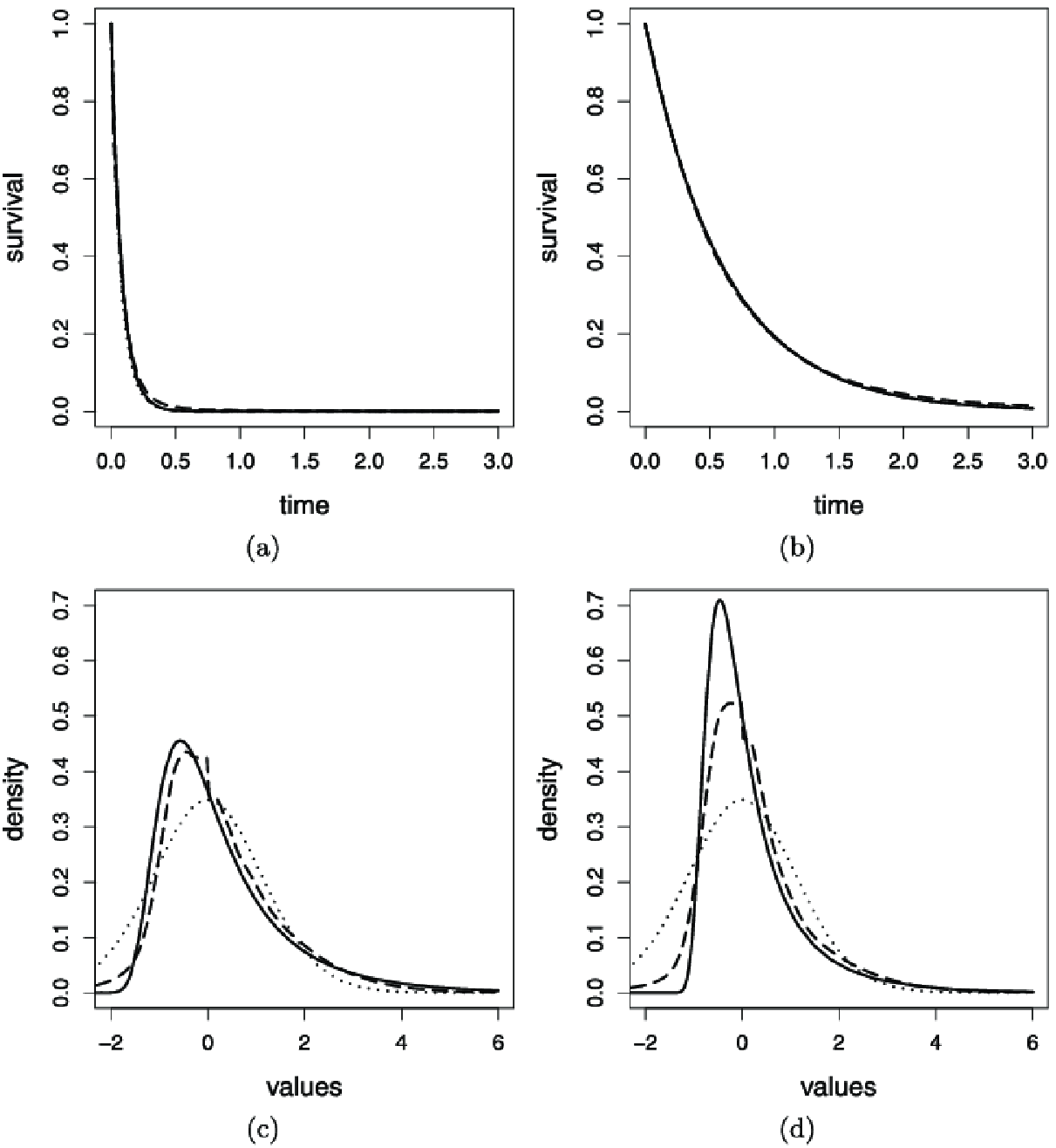}

\caption{Simulated data---Scenario \textup{II}: Mean, across simulations, of
the posterior mean of the survival and frailty density functions under
the proposed model. Panels \textup{(a)} and \textup{(b)} show the results for the
survival functions. Panels \textup{(c)} and \textup{(d)} show the results for the frailty
densities. Panels \textup{(a)} and \textup{(c)} show the results for covariate values
$(2,1,0.5)$. Panels \textup{(b)} and \textup{(d)} show the results for covariate values
$(0,1,1.5)$. The true curves are represented by continuous lines. The
results under the proposed model are represented by dashed lines. The
results under the exchangeable Gaussian frailty model are represented
by dotted lines. In panels \textup{(a)} and \textup{(b)} the results obtained under the
PSF approach are represented by dot-dashed lines.}
\label{case2}
\end{figure}

The average of the estimated frailty distributions and survival
functions across simulated data sets for some specific covariate values
are presented in Figure~\ref{case1} for {Scenario I} and in Figure~\ref
{case2} for {Scenario II}. The results in {Scenario I} reveal that the
proposed model roughly captures the modal behavior of the
covariate-dependent frailty distributions. Although not perfect, the
proposed model performs remarkably well given that only $n=100$
imperfectly-observed observations were generated for each data set. The
situation is much less favorable for the GF approach, which fails to
correctly capture the shape of the frailty distributions, leading to
poor estimated survival functions. This behavior is likely driving the
underestimation of survival noted in the SEER analysis. As expected,
the PSF approach also suffers from bad prediction since the underlying
assumption for frailty distribution is violated. The results in
{Scenario II} show that the proposed model is still able to capture the
frailty distributional shape even when the data were truly generated
from the PSF model. Regarding the estimated survival curves, the
results suggest that essentially no differences among the three methods
are observed; all estimated functions are close to the truth,
indicating that there is little price to be paid when using the
proposed model for the clustered survival data that were truly
generated from the PSF model.

%

%

The results of the comparison of the estimated survival curves in terms
of ISE are presented in Table~\ref{ttwo}, where the Monte Carlo mean
and standard deviations for the ISE for two different predictor values
are given. The results under {Scenario I} show a close agreement with
the observed for the regression coefficients; the proposed model
substantially outperforms the other two methods in terms of smaller
means and standard deviations of the ISE. Even under {Scenario II}, the
proposed model still provides almost the same results as the PSF model
in terms of ISE.

%
\begin{table}
\tabcolsep=0pt
\caption{Simulated data---Scenario \textup{II}: Monte Carlo mean (Monte Carlo
standard deviation) for the ISE of the survival function for two
different predictor values. The results for the different approaches
under both simulation scenarios are presented. The numbers correspond
to $10^3 $ times the original values}\label{ttwo}
\begin{tabular*}{\tablewidth}{@{\extracolsep{\fill}}@{}lcccc@{}}
\hline
\textbf{Scenario} & $\bolds{(w_1, w_2, x)}$ & \textbf{Proposed model} & \textbf{GF model} & \textbf{PSF model}\\
\hline
\phantom{I}I & $(2,1,-2)$ & 2.02 (2.48) & 4.37 (3.46) & 6.28 (3.49) \\
 & $(0,1,2)$ & 1.94 (2.53) & 10.5 (6.86) & 14.3 (10.9)
\\[3pt]
II & $(2,1,0.5)$ & 3.17 (4.66) & 3.13 (3.33) & 2.19 (2.26) \\
 & $(0,1,1.5)$ & 0.96 (1.18) & 0.89 (1.22) & 0.83 (1.10) \\
\hline
\end{tabular*}
\end{table}

In Appendix~D of the supplementary material [\citet{Zhouetal2014}],
additional simulation results are presented which show that, under
{Scenario I}, for larger sample sizes better estimates of the frailty
distributions are obtained and that the approach is not affected by the
choice of $J$ in the specification of the LDTFP model. For further
comparison, we also fitted the exchangeable mixture of Polya trees
(MPT) [\citet{Hanson2006}] frailty Cox's model using the function \texttt
{PTglmm} available in \texttt{DPpackage} [\citet{Jaraetal2011}] under
{Scenario I}, in which the results show that our approach outperforms
the MPT, and considered a third scenario favorable to the GP approach,
where the results show that our method pays little price for the extra
generality when using the proposed model when normality and
exchangeability are valid assumptions. Overall, the proposed approach
provides a flexible way to capture the heterogeneity in the frailty
distribution, provides superior prediction, and yields an essential
improvement for the estimation of population effects, especially when
the intra-cluster correlation (or variability in frailties) is
relatively large. When the frailty variances are small across clusters,
the proposed approach is still recommended due to its flexibility.

\section{Concluding remarks}\label{sec5}
Very limited work has been done on covariate-adjusted frailty survival
models for clustered time-to-event data. Liu,\break Kalbfleisch and
Schaubel
(\citeyear{Liuetal2011}) proposed a
stratified Cox model with positive stable frailties, where the shape
parameter of the frailty distribution is allowed to depend on
cluster-level covariates. However, they essentially fitted a marginal
Cox model, and then utilized the positive stable assumption and some
imposed constraints to estimate the parameters in their proposed model.
The model proposed in this paper cleanly separates population-level
effects from the cluster-level effects, which determine the shape of
the frailty distribution. Frailty density shape is modeled using a
tractable median-zero LDTFP prior. Other nonparametric density
regression approaches could also be considered; however, model
identifiability requires a location constraint such as mean-zero or
median-zero. The proposed model provides a natural generalization of
the conventional PH model with parametric or nonparametric exchangeable
frailties, and accommodates frailty distribution ``evolution'' over
cluster-level covariates providing superior prediction, as shown in our
simulation studies. When data are truly generated according to an
exchangeable Gaussian frailty PH model or the model of \citet
{Liuetal2011}, our model does about the same as the underlying true
model in terms of fixed effects and/or marginal survival estimations.
We illustrate the usefulness of the proposed model with an analysis of
a subset of the Iowa SEER breast cancer data, and demonstrate that
higher degree of ruralness corresponds to a more bimodal frailty
distributional shape with larger variance. In general, the proposed
model is more flexible than currently existing frailty PH models,
leading to more robust inferences, and thus is recommended. One
drawback of the proposed model is that, as currently fit in R,
obtaining inference takes longer.

For ease of computation, we have assumed a piecewise constant structure
for the baseline hazard function $\lambda_0(t)$ and taken the
independent normal prior distributions for $\log(\lambda_k)$'s, so that
the baseline hazard heights $\lambda_k$ and covariate effects $\bxi$
can be updated simultaneously. The use of empirically-derived
cut-points has permeated much of the Bayesian survival literature for
over a decade. Use of Breslow's baseline estimate coupled with the GF
approach led to a greatly increased LPML over the empirical approach.
An obvious extension of our current work is to employ a smoothed
baseline, for example, using penalized B-splines [\citet
{Hennerfeindetal2006}], the piecewise exponential with random
cut-points [\citet{SahuDey2004}], MPT [\citet{Hanson2006}], etc. Any of
these approaches could improve model fit and prediction, but cannot
currently be fitted in the \texttt{R} software. We are currently
working on extending the methodology in this paper to other survival
models and smoothed baselines.

\section*{Acknowledgments}
We thank the referees and editors for numerous insightful suggestions
which greatly enhanced the readability of this paper.


\begin{supplement}\label{suppA}
\stitle{Supplement to ``Modeling county-level breast cancer survival
data using a covariate-adjusted frailty proportional hazards model''\\}
\slink[doi]{10.1214/14-AOAS793SUPP} 
\sdatatype{.pdf}
\sfilename{aoas793\_supp.pdf}
\sdescription{In this online supplemental article we provide (A)
technical details on the mixture of linear dependent tailfree
processes, (B) a detailed description of the MCMC algorithm, (C) sample
R code to analyze the SEER data, (D) additional simulation studies and
(E) additional analysis of the SEER data.}
\end{supplement}

%

\printaddresses
\end{document}